\documentclass{emulateapj}

\shorttitle{$\Gamma-L_{\rm \gamma,iso}(E_{\rm \gamma,iso})$ Correlations of GRBs and Their Interpretation}
\shortauthors{L\"u et al.}

\begin{document}
%---------------------------Title-----------------------------
\title{Lorentz factor - isotropic luminosity/energy correlations of GRBs and their interpretation}

\author{Jing L\"u\altaffilmark{1}, Yuan-Chuan Zou\altaffilmark{1}, Wei-Hua Lei\altaffilmark{1,2}, {Bing Zhang\altaffilmark{2}}, Qing-Wen Wu\altaffilmark{1}, Ding-Xiong Wang\altaffilmark{1}, {En-Wei Liang\altaffilmark{3,4}}, and Hou-Jun L\"u\altaffilmark{2}}

\altaffiltext{1}{School of Physics, Huazhong University of Science and
Technology, Wuhan, 430074, China. Email: zouyc@hust.edu.cn (YCZ), leiwh@hust.edu.cn (WHL)}
\altaffiltext{2}{Department of Physics and Astronomy, University of Nevada Las Vegas, 4505 Maryland Parkway, Box 454002, Las Vegas, NV 89154-4002, USA. Email: zhang@physics.unlv.edu (BZ)}
\altaffiltext{3}{Department of Physics, Guangxi University, Nanning 530004, China}
\altaffiltext{4}{GXU-NAOC Center for Astrophysics and Space Sciences, Nanning, Guangxi 530004, China}

%-------------------------Abstract----------------------------
\begin{abstract}
The bulk Lorentz factor of the gamma-ray burst (GRB) ejecta ($\Gamma_{\rm 0}$) is a key parameter to understand the GRB physics. Liang et al. have discovered a correlation between $\Gamma_0$ and isotropic $\gamma$-ray energy: $\Gamma_{\rm 0}\propto E_{\rm \gamma,iso,52}^{0.25}$.
By including more GRBs with updated data and more methods to derive $\Gamma_0$, we confirm this correlation
and obtain $\Gamma_{\rm 0} \simeq 91 E_{\rm \gamma,iso,52}^{0.29}$. Evaluating the mean isotropic $\gamma$-ray
luminosities $L_{\rm \gamma,iso}$ of the GRBs in the same sample, we discover an even tighter correlation
$\Gamma_{\rm 0} \simeq 249 L_{\rm \gamma,iso,52}^{0.30}$. We propose an interpretation to this later correlation.
Invoking a neutrino-cooled hyperaccretion disk around a stellar mass black hole as the central engine of GRBs,
we derive jet luminosity powered by neutrino annihilation and baryon loading from a neutrino-driven wind. Applying beaming correction, we finally derive $\Gamma_{\rm 0} \propto L_{\gamma, \rm iso}^{0.22}$, which is consistent with the data. This suggests that the central engine of long GRBs is likely a stellar mass black hole surrounded by a hyper-accreting disk.
\end{abstract}


%-------------------------Introduction----------------------------
\section{Introduction}
\label{intro}

Gamma-ray bursts (GRBs) are among the most powerful explosions in the universe \citep{piran04,meszaros06,zhang07}. It is well known that GRBs are produced by relativistic outflows. The bulk Lorentz factor during the prompt GRB emission phase ($\Gamma_{\rm 0}$, also called ``initial'' Lorentz factor to be differentiated from the decaying Lorentz factor during the afterglow phase) is a very important parameter to understand the physics of GRBs. There have been several methods to infer $\Gamma_{\rm 0}$: (1) Taking the peak time of the early afterglow light curve as the deceleration time of the external forward shock, one can estimate $\Gamma_{\rm 0}$, which is twice of the Lorentz factor at the deceleration time \citep{b2}. A non-detection of such an afterglow peak time due to a late response time or contamination of other emission components can lead to a lower limit of $\Gamma_0$ \citep[e.g.][]{Zhang2006}. (2) The ``compactness problem" constraint \citep{Piran1999}, i.e., the requirement that GRBs are optically thin to two photon pair production also yields a lower limit on $\Gamma_{\rm 0}$ \citep{Lithwick2001,Gupta2008,Abdo2009a,Abdo2009b,Abdo2009c}. (3) During the prompt emission phase, the external shock is already growing \citep[e.g.][]{Maxham09}. An upper limit of $\Gamma_{\rm 0}$ can be derived from the data based on the requirement that the external shock emission is not bright enough during the prompt emission phase \citep{Zou2010}.

Constraining $\Gamma_0$ of GRBs and studying their statistical properties is essential to constrain the physical origin of GRB prompt emission \citep{Liang2010}. In particular, different theoretical models demand different correlations between $\Gamma_0$ and $E_{\rm \gamma,iso}$ or $L_{\rm \gamma,iso}$ \citep{Zhang2002} in order to account for the observed $E_{\rm p}-E_{\rm \gamma,iso}$ correlation \citep{Amati2002}. By constraining $\Gamma_0$ of about 20 GRBs that show the deceleration feature in the early afterglow lightcurves, \cite{Liang2010} discovered a tight correlation between $\Gamma_{\rm 0}$ and $E_{\rm \gamma,iso}$, i.e. $\Gamma_{\rm 0} \simeq 182 (E_{\rm \gamma,iso}/10^{52}{\rm erg})^{0.25}$. Using a different method to derive $\Gamma_0$, \cite{Ghirlanda2011} confirmed a correlation between $\Gamma_0$ and $E_{\rm \gamma,iso}$, but with a different power index\footnote{The Ghirlanda et al. (2011) method applies the Blandford-Mckee (BM) self-similar deceleration solution \citep{bm76} and extrapolates it backwards to derive $\Gamma_0$. However, around the deceleration stage, the dynamics has not entered the BM self-semilar solution yet. Also the intersection of the two asymptotic power law phases (as adopted by Ghirlanda et al. 2011) may not correspond to the observed peak time of afterglow light curve. We regard the Ghirlanda et al. (2011) method not more precise than the conventional method, and still adopt the conventional method to derive $\Gamma_0$ in this paper.}.

In this paper, we work on an expanded sample and apply more methods to constrain $\Gamma_0$ for about 50 GRBs. We test and confirm the $\Gamma_0-E_{\rm \gamma,iso}$ correlation discovered by \cite{Liang2010}, and also investigate a $\Gamma_0 - L_{\rm \gamma,iso}$ correlation, where $L_{\rm \gamma,iso}$ is the mean luminosity of the burst. Since the $\Gamma_0-E_{\rm \gamma,iso}$ correlation was not interpreted in the previous work \citep{Liang2010}, we also attempt to propose an interpretation in this paper. We derive jet luminosity and baryon loading from a black hole - neutrino-cooling-dominated-flow (NDAF) disk central engine model, and find that this central engine model can naturally account for the $\Gamma_0 - L_{\rm \gamma,iso}$ correlation discovered in this paper, and hence, can also interpret the $\Gamma_0-E_{\rm \gamma,iso}$ correlation of \cite{Liang2010}.

We arrange this paper as follows. In Section 2, the methods of $\Gamma_{\rm 0}$ derivations based on three methods
are summarized. We then apply the methods to the available GRBs on which these methods can be used, and present the $\Gamma_0-E_{\rm \gamma,iso}$ and $\Gamma_0 - L_{\rm \gamma,iso}$ correlations in Section 3. In Section 4, a physical interpretation to the $\Gamma_0 - L_{\rm \gamma,iso}$ correlation is presented. Our results are summarized in Section 5 with some discussion.

\section{Methods of Constraining $\Gamma_0$}
\label{main}

We apply three methods to constrain $\Gamma_{\rm 0}$, namely, (A) the afterglow onset method \citep{b2}, (B) pair opacity constraint method \citep{Lithwick2001}, and (C) early external forward emission method \citep{Zou2010}.

Method A is the most common method, which uses the peak of the early afterglow light curve to determine the deceleration time of the external forward shock. In the so-called ``thin shell'' regime, the initial Lorentz factor $\Gamma_0$ is twice of the Lorentz factor at the deceleration time. For a constant density medium, one has
\begin{equation}
   \Gamma_{\rm 0}\simeq1.4\left[\frac{3E_{\rm \gamma,iso}(1+z)^{3}}{32{\pi}nm_{\rm p}c^{5}{\eta}t_{\rm peak}^{3}}\right]^{1/8},
   \label{eq:dec}
\end{equation}
where $n$ is the medium number density, $m_{\rm p}$ is the proton rest mass, $\eta$ is the ratio between the isotropic gamma-ray energy and the isotropic blast wave kinetic energy, and $t_{\rm peak}$ is peak time of the afterglow, which is also taken as the deceleration time. The derived $\Gamma_0$ is rather insensitive to $n$ and $\eta$, but mildly depends on $t_{\rm peak}$ (-3/8 power). If the peak time is not detected, $t_{\rm peak}$ is regarded to be prior to the earliest afterglow observing time (i.e., $t_{\rm obs}>t_{\rm peak}$). This gives a lower limit on $\Gamma_0$. Notice that in Eq.(\ref{eq:dec}), we have taken the coefficient as 1.4 rather than the commonly used 2. This more precise factor comes from two factors: First, the deceleration radius is defined by the condition $M=M_0/ \Gamma_{\rm dec}$ rather than $M=M_0/\Gamma_0$ (where $M$ is the shocked ISM mass, and $M_0$ is the original mass of the ejecta), since at this radius, the shocked ISM and the ejecta have the same inertia. Second, instead of adopting $r_{\rm dec} \simeq 2 \Gamma_{\rm dec}^2 c t_{\rm dec}$, we apply a differential form ${\rm d}r \simeq 2 \Gamma^2 c {\rm d}t$, and numerically integrate it from $t=0$ to $t=t_{\rm dec}$ to get $r_{\rm dec} = 4.4 \Gamma_{\rm dec}^2 c t_{\rm dec}$. Here $r_{\rm dec}$ is the deceleration radius and $t_{\rm dec}$ is the deceleration time, which also corresponds to the peak time $t_{\rm peak}$.

Method B requires that observed high energy $\gamma$-rays (e.g. those in the GeV range) are optically thin to
electron-positron pair production with softer target photons in the emission region. This yields a lower limit on the Lorentz factor of the emitting region \citep{Lithwick2001}\footnote{This method makes the assumption that the GRB emission radius $R_{\rm GRB}$ is related to $\Gamma_0$ via $R_{\rm GRB} \simeq \Gamma_0^2 c \delta T$. Some GRB prompt emission models do not satisfy such a condition \citep[e.g.][]{Narayan09,Zhang2011}. The lower limit of $\Gamma_0$ cannot be uniquely derived, since the cutoff energy is a function of both $\Gamma_0$ and $R_{\rm GRB}$ \citep{Gupta2008}.}. The lower limit can be obtained by requiring that the observed highest energy photons with energy $E_{\max}$ have an optical depth smaller than unity:
\begin{equation}
   \Gamma_{\rm 0}>\hat{\tau}^{\frac{1}{2(\beta+2)}}\left(\frac{E_{\rm max}}{m_{\rm e} c^2}\right)^{\frac{\beta-1}{2\beta+2}}(1+z)^{\frac{\beta-1}{\beta+1}},
   \label{eq:compact}
\end{equation}
and
\begin{equation}
\hat{\tau}=2.1\times10^{11} \left[\frac{(D/7{\rm Gpc})^2(0.511)^{-\beta+1}f_{\rm 1}}{(\delta T/0.1{\rm s})(\beta-1)}\right],
\label{eq:tau}
\end{equation}
where $\beta$ is the photon spectral index in the MeV band, with a typical value between 2 and 3, $D$ is the luminosity distance, $\delta T$ is the minimum variability time scale of the prompt emission, and $f_{\rm 1}$ is the observed number of photons per second per ${\rm cm}^2$ per MeV at the energy of 1 MeV \citep{Lithwick2001}. We notice that there are a few bursts whose $\Gamma_0$'s constrained using this method are inconsistent with those derived from other two methods. Instead, we apply a modified version of Method B, which assumes the high energy emission and the prompt MeV emission are from two different emitting regions \citep{Zou2011,Zhao2011}.

Method C considers the quiescent periods between the prompt emission pulses, in which the signal of external shock has to go down the instrument thresholds. This would place an upper limit on $\Gamma_{\rm 0}$ \citep{Zou2010}. The constraint of $\Gamma_{\rm 0}$ for a uniform density medium is
\begin{eqnarray}
   \Gamma_0<  340(1+z)^{\frac{1}{4}}{f_{\rm {\nu,lim,-28}}}^{\frac{1}{9}}D_{28}^{\frac{2}{9}}n_{0}^{-{\frac{1}{8}}}\epsilon_{\rm e,-{\frac{1}{2}}}^{-{\frac{1}{6}}}\epsilon_{\rm B,-1}^{-{\frac{1}{72}}}\nu_{20}^{\frac{5}{36}}t_{\oplus}^{-{\frac{2}{9}}}(1+Y)^{\frac{1}{9}},
   \label{eq:early}
\end{eqnarray}
where $f_{\rm {\nu,lim}}\sim 10^{-28} {\rm erg \, cm^{-2} s^{-1} Hz^{-1}}$ is the limiting flux density of the observing instrument, $Y$ is the Compton parameter for synchrotron self-Compton scattering, $\epsilon_{\rm e}$ is the equipartition factor for internal energy density of electrons, $\epsilon_{\rm B}$ is the equipartition factor for the magnetic energy density, and $t_{\oplus}$ is the first quiescent time in the observer's frame, and $\nu$ is the observing frequency. In this paper, we take the conventional notation $Q = Q_k \times 10^k$ if not specified.

\section{Sample Selection and Correlations}

Using the methods above, we can constrain $\Gamma_0$ for the bursts with enough observational data. The parameters of 51 GRBs in our sample are presented in Table 1, which include spectroscopically confirmed redshift ($z$), burst duration ($T_{\rm 90}$), derived initial Lorentz factor $\Gamma_{\rm 0}$, isotropic $\gamma$-ray energy ($E_{\rm \gamma,iso}$), and isotropic mean $\gamma$-ray luminosity ($L_{\rm \gamma,iso} \equiv (1+z) E_{\rm \gamma,iso}/T_{\rm 90}$). Within the sample, 38 GRBs have $\Gamma_{\rm 0}$ calculated using Method A (Refs a, b and d in Table 1). As methods B and C can only get a range for the derived Lorentz factor, the fit for the relations of $\Gamma_{\rm 0}-E_{\rm \gamma,iso}$ and $\Gamma_{\rm 0}-L_{\rm \gamma,iso}$ are from these 38 GRBs only.

With the data listed in Table 1, a correlation analysis between $\log\Gamma_{\rm 0}$ and $\log{L_{\rm \gamma,iso}}$ data set yields a Pearsons correlation coefficient with $\zeta=0.79$, which is tighter  than the $\log\Gamma_0 - \log{E_{\rm \gamma,iso}}$ correlation with $\zeta=0.67$. We plot $\Gamma_{\rm 0}$ versus $E_{\rm \gamma,iso}$ and $L_{\rm \gamma,iso}$ in Fig. 1 and Fig. 2, respectively. Visibly one can see a strong correlation in both plots. The best fitting results are:
\begin{equation}
   \log{\Gamma_{\rm 0}}=(1.96\pm0.002)+(0.29\pm0.002)\log{E_{\rm \gamma,iso,52}}
\end{equation}
with $\zeta=0.67$, and
\begin{equation}
   \log{\Gamma_{\rm 0}}=(2.40\pm0.002)+(0.30\pm0.002)\log{L_{\rm \gamma,iso,52}}
\end{equation}
with $\zeta=0.79$.

These correlations can be translated to
\begin{equation}
   \Gamma_{\rm 0}\simeq 91E_{\rm \gamma,iso,52}^{0.29},
\end{equation}
and
\begin{equation}
   \Gamma_{\rm 0}\simeq 249L_{\rm \gamma,iso,52}^{0.30}.
\label{eq:L}
\end{equation}

It can be seen that the $\Gamma_0 - E_{\rm \gamma,iso}$ correlation discovered by \citet{Liang2010} is confirmed. The smaller coefficient (91 instead of 182) is mainly caused by the smaller (but more precise) factor 1.4 (rather than 2) in Eq.(\ref{eq:dec}). We also found a tighter $\Gamma_0 - L_{\rm \gamma,iso}$ correlation, suggesting that it may be more intrinsic than the $\Gamma_0 - E_{\rm \gamma,iso}$ correlation. As $\Gamma_0$, $E_{\rm \gamma,iso}$ and $L_{\rm \gamma,iso}$  are all $z$-dependent quantities, there might be a selection effect involved so that the correlation may not be intrinsic \citep{bkb09}. In order to test this possibility, we study the $\Gamma_0 - L_{\rm \gamma,iso}$ relation with the following procedure: 1. We randomly produce a set of redshifts according to the GRB $z$-distribution given by \cite{wp10}; 2. assign these random artificial redshifts to the bursts to replace the observed ones; 3. calculate the $\Gamma_0$ and $L_{\rm \gamma,iso}$ according to the artificial redshifts; 4. calculate the correlation coefficient $\zeta$ of $\log \Gamma_0-\log L_{\rm \gamma,iso}$ correlation for each realization; 5. redo step 1 through 4 10000 times, and get a distribution of correlation coefficient; 6. compare the most probable coefficient with the coefficient generated from the real data. The most probable coefficient from our simulations is 0.63, which is clearly smaller than the one derived from the real data, $\zeta=0.79$. This means that the $\Gamma_0 - L_{\rm \gamma,iso}$ relation is likely intrinsic, not caused by a selection effect from $z$-dependence parameters.

We notice two outliers to both correlations: GRB060614 and GRB080129, whose $\Gamma_0$'s are derived using Method A from a late optical bump, which lead to $\Gamma_0 < 100$ for both cases. It is possible that these bumps are caused by other mechanisms \citep[e.g. energy injection,][]{Xu2009}. If this is the case, the derived $\Gamma_0$ for the two bursts can be regarded as lower limits.

\section{Theoretical Interpretation}
\label{model}

The most popular model of GRB central engine invokes a stellar mass black hole surrounded by a hyper-accreting disk \citep[e.g.][]{Popham1999, Narayan2001, Di2002, Kohri2002, Gu2006, ChenWX2007, Janiuk2007, Lei2009}. In the inner region of such a hyperaccretion disk a large amount of energetic neutrinos are emitted, carrying away the viscous dissipation energy of the accreted gas. If the accretion rate is not too low, neutrino annihilation ($\nu \bar{\nu} \rightarrow e^{+} e^{-}$) can launch a relativistic jet powerful enough to account for the GRB.

For a system with black hole mass $M$ and spin $a_*$, the neutrino annihilation power $\dot{E}_{\nu\bar{\nu}}$ from the hyperaccretion disk depends on the accretion rate $\dot{M}$ (for $\dot{M}_{\rm ign}< \dot{M}<\dot{M}_{\rm trap} $) as \citep{Zalamea2011},
\begin{equation}
\dot{E}_{\nu \bar{\nu}} \simeq 1.1 \times 10^{52} x_{\rm ms}^{-4.8} M_3^{-3/2} \dot{m}^{9/4} {\rm erg \ s^{-1}},
\label{eq:Evv}
\end{equation}
where $M_3=M/3M_{\sun}$, $\dot{m}=\dot{M}/M_{\sun} {\rm s^{-1}}$, $x_{\rm ms} \equiv r_{\rm ms}(a_*)/r_g$, and $r_g=2 G M/c^2$. Here $r_{\rm ms}$ is the radius of the marginally stable orbit, which is a function of the black hole spin $a_*$ \citep{Page1974}. We have $x_{\rm ms} =0.97$ for $a_*=0.95$. The two critical accretion rates $\dot{M}_{\rm ign}$ and $\dot{M}_{\rm trap}$ are defined in \citet{Zalamea2011}. If $\dot{M} <\dot{M}_{\rm ign}$, the disc temperature is not high enough to ignite neutrino emitting reactions. If $\dot{M} > \dot{M}_{\rm trap}$ , the emitted neutrinos become trapped in the disc and advected into the black hole. For the disk with viscosity $\alpha=0.1$, we find $\dot{M}_{\rm ign} = 0.071 M_{\sun} {\rm s^{-1}}$ and $\dot{M}_{\rm trap}=9.3 M_{\sun} {\rm s^{-1}}$ for $a_*= 0$, and $\dot{M}_{\rm ign} = 0.021 M_{\sun} {\rm s^{-1}}$ and $\dot{M}_{\rm trap}=1.8 M_{\sun} {\rm s^{-1}}$ for $a_*= 0.95$.

Most neutrino annihilation energy is converted into kinetic energy of baryons after acceleration, and the jet reaches a Lorentz factor
\begin{equation}
\Gamma_0 \simeq  \frac{\dot{E}_{\nu \bar{\nu}}}{\dot{M}_{\nu} c^2}
\label{eq:Gamma}
\end{equation}
where $\dot{M}_{\nu}$ is the neutrino-driven mass loss rate from the disk. The mass loss rate $\dot{M}_{\nu}$ is related to the total neutrino power $\dot{E}_\nu$ through \citep{Metzger2008}
\begin{equation}
\dot{M}_{\nu} \simeq 10^{-6} \dot{E}_{\nu,52}^{5/3} \langle\epsilon_{10}^2\rangle^{5/3} r_6^{5/3} M_3^{-2}(h/r)^{-1} M_{\sun} {\rm s}^{-1}
\label{eq:Mv}
\end{equation}
where $r_6=r/10^6 {\rm cm}$, $\dot{E}_{\nu,52}=\dot{E}_{\nu}/10^{52} {\rm erg \ s^{-1} }$, $\epsilon_{\nu} = \epsilon_{10} \times 10 {\rm MeV}$ is the mean energy of neutrinos, and $h$ is the half-thickness of disk. For $a_*=0.95$, the total neutrino power from the disk is $\dot{E}_{\nu} \simeq 0.15 \dot{M}c^2$  \citep{ChenWX2007}.

For a neutrino dominated accretion flow (NDAF), both $\epsilon_{\nu}$ (which is a function of disk temperature) and $h$ are independent of the accretion rate $\dot{m}$. This result can be checked with the analytical solution of hyper-accreting disk obtained by \citet{Popham1999} (i.e., their equations (5.3)and (5.4)). So, based on Eq.(\ref{eq:Mv}), the mass loss rate $\dot{M}_{\nu}$ is just related to $\dot{m}$ as $\dot{M}_{\nu} \propto \dot{E}_{\nu}^{5/3} \propto \dot{m}^{5/3}$. Combining this dependence with Eq.(\ref{eq:Evv}), one drives $\dot{M}_{\nu} \propto \dot{m}^{5/3} \propto \dot{E}_{\nu \bar{\nu}}^{20/27}$. And then inserting it to Eq.(\ref{eq:Gamma}), we therefore obtains $\Gamma_0 \propto \dot{E}_{\nu \bar{\nu}}/\dot{M}_{\nu} \propto \dot{E}_{\nu \bar{\nu}}^{7/27}$.

The relativistic jet with Lorentz factor $\Gamma_0$ will dissipate its kinetic energy via internal shocks with efficiency $\eta$ and produce gamma-ray emission, i.e., $L_{\gamma} \simeq \eta_{\gamma} \dot{E}_{\nu\bar{\nu}}$. Assuming a constant $\eta_\gamma$ for all GRBs, one can get $\Gamma \propto L_{\rm \gamma}^{7/27}$. In order to connect $L_\gamma$ and $L_{\rm \gamma,iso}$, one needs to further take into account the beaming correction, i.e. $L_\gamma = f_b L_{\rm \gamma,iso}$. One then gets $L_{\rm \gamma,iso} =f_b^{-1} L_{\gamma}=f_b^{-1} \eta_{\gamma} \dot{E}_{\nu \bar{\nu}} \propto f_b^{-1} \dot{E}_{\nu \bar{\nu}}$, where $f_b \ll 1$ is the beaming factor.

The general dependence of $f_b$ on the properties of central engine is unknown. However, one can gain insight directly from observations. Following Amati et al. (2002, 2006, 2008), the relationship between the isotropic equivalent energy radiated during the prompt phase ($E_{\gamma,{\rm iso}}$) and the rest-frame peak energy in the GRB spectrum ($E_p^{\prime}$ ) is $E_p^{\prime} \propto E_{\gamma,{\rm iso}}^{0.57}$. By combining it with the Girlanda relation $E_{\gamma} \propto (E_p^{\prime} )^{3/2} $ (Ghirlanda et al. 2004), where the beaming-corrected energy $E_{\gamma}=f_b E_{\gamma,{\rm iso}}$, we obtain the relation between $f_b$ and $E_{\gamma,{\rm iso}}$ as $f_b \propto E_{\gamma,{\rm iso}}^{-0.145}$. Since $L_{\gamma,{\rm iso}} \propto E_{\gamma,{\rm iso}}$, we get $f_b \propto L_{\gamma,{\rm iso}}^{-0.145}$. One can see that $f_b$ is very insensitive to $L_{\rm \gamma,iso}$ and $E_{\rm \gamma,iso}$.

Now we can obtain the relation between Lorentz factor $\Gamma_0$ and the isotropic luminosity $L_{\rm \gamma,iso}$ based on the above scalings, i.e.
\begin{eqnarray}
\Gamma_0 \propto \dot{E}_{\nu \bar{\nu}}^{7/27} \propto (f_b L_{\rm \gamma,iso})^{7/27} \propto L_{\rm \gamma,iso}^{0.22}~.
\end{eqnarray}
In view of the large scatter of the applied empirical Amati- and Ghirlanda-correlations, we regard that this theoretically motivated correlation agrees with the statistical correlation (\ref{eq:L}).

\section{Conclusions and Discussion}
\label{conclusion}

By including more recent GRBs and by engaging more methods to constrain $\Gamma_0$, we have critically re-analyzed the statistical correlation between $\Gamma_0$ and $E_{\rm iso}$ \citep{Liang2010}. We confirmed the correlation and found $\Gamma_0 \simeq 91 E_{\rm \gamma, iso, 52}^{0.29}$. Furthermore, we found an even tighter correlation between $\Gamma_0$ and the mean isotropic $\gamma$-ray luminosity, which reads $\Gamma_0 \simeq 249 L_{\rm \gamma, iso, 52}^{0.30}$.

We also proposed an interpretation to the $\Gamma_0 \sim L_{\rm iso}^{0.30}$ correlation within the framework of a
black hole - NDAF disk GRB central engine model. By invoking a neutrino-annihilation powered jet and by calculating
baryon loading from a neutrino-driven wind, we get a $\Gamma_0 \propto L_{\rm \gamma}^{7/27}$ correlation. Further
considering the beaming factor $f_b$, which is insensitive to $L_{\rm \gamma,iso}$ as evidenced from the empirical
Amati and Ghirlanda correlations, we finally derived $\Gamma_0 \sim L_{\rm \gamma,iso}^{0.22}$. In view of the large scatter of various correlation, we regard that this model prediction is well consistent with the observed $\Gamma_0-L_{\rm \gamma,iso}$ correlation.

The existence of the $\Gamma_0-L_{\rm \gamma,iso}$ and  $\Gamma_0-E_{\rm \gamma,iso}$ correlations and the success of interpreting them within the black hole - NDAF central engine model hint that the GRB central engine is likely a hyper-accreting black hole. The interpretation invokes a neutrino-annihilation-powered jet, which is justified for a reasonably high accretion rate and a not very rapid black hole spin (W.-H. Lei \& B. Zhang 2011, in preparation). On the other hand, recently arguments have been raised to support a magnetically dominated jet from GRBs \citep[e.g.][]{Zhang2009,Fan2010,Zhang2011}. Studies of the black hole central engine models also suggest that magnetic fields play an important role \citep[e.g.][]{Lei2009}. The baryon loading process in a magnetically dominated jet is more complicated, and has not been studied carefully in the literature. Whether the $\Gamma_0-L_{\rm \gamma,iso}$ correlation can be still interpreted in a magnetized black hole central engine model \citep[e.g.][]{Blandford77,Meszaros97,Wang02,Yuan2012} is subject to further investigations.

Since only one short GRB (090510) is included in our sample, our correlations and interpretation apply to long GRBs only.

Recently, \cite{Wu2011} discovered an intriguing universal correlation between synchrotron luminosity and Doppler factor for GRBs and blazars. Our interpretation cannot be extended to blazars, since the accretion rate inferred from blazars are not in the NDAF regime. If indeed the two phenomenon share the same physics, then the correlation may stem from a more profound physical origin, which is beyond the scope of this paper.

\acknowledgments

We thank T. Piran, Y. C. Ye, and H. Gao for helpful discussion. This work is supported by NSF under Grant No. AST-0908362, NASA under Grant No. NNX10AD48G, National Natural Science Foundation of China (grants 11173011, 11143001, 11133005, 11103003, 11003004, 11025313, 10873002, 10873005 and 10703002), National Basic Research Program (``973'' Program) of China grant 2009CB824800, Fundamental Research Funds for the Central Universities (HUST: 2011TS159), Guangxi Natural Science Foundation (2010GXNSFC013011 and 2011-135) and Guangxi SHI-BAI-QIAN project (grant 2007201). WHL acknowledges a Fellowship from China Scholarship Program for support.

%-----------------------------------------------------------------------------

%--------------------- Figure --------------------------
\clearpage
\begin{figure}
\begin{center}
\includegraphics[width=\textwidth]{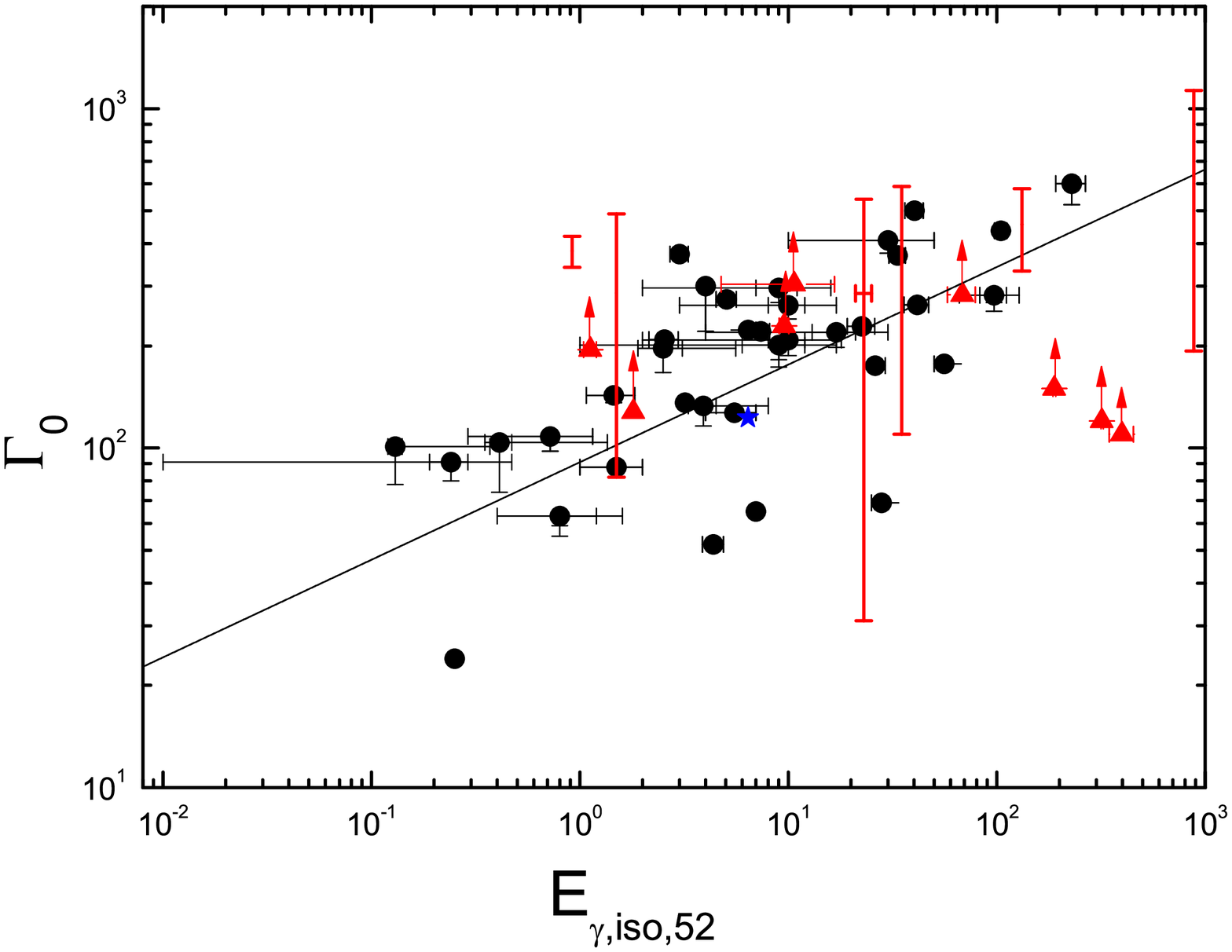}
 \caption{The plot of the derived initial Lorentz factor $\Gamma_{\rm 0}$ vs. the isotropic equivalent $\gamma$-ray energy $E_{\rm \gamma,iso}$. The solid line, $\Gamma_{\rm 0} \simeq 91 E_{\rm \gamma,iso,52}^{0.29}$, is the best fit to the derived values, which are in solid circles.
The Pearson's correlation coefficient is $\zeta=0.67$.
 The triangles are the bursts having only lower limits and the range segment are the bursts having upper and lower limits listed in Table 1, which are not included in the fitting process. The star is the only short burst GRB 090510.}
\end{center}
\label{fig:GE}
\end{figure}

\begin{figure}
\begin{center}
{\includegraphics[width=\textwidth]{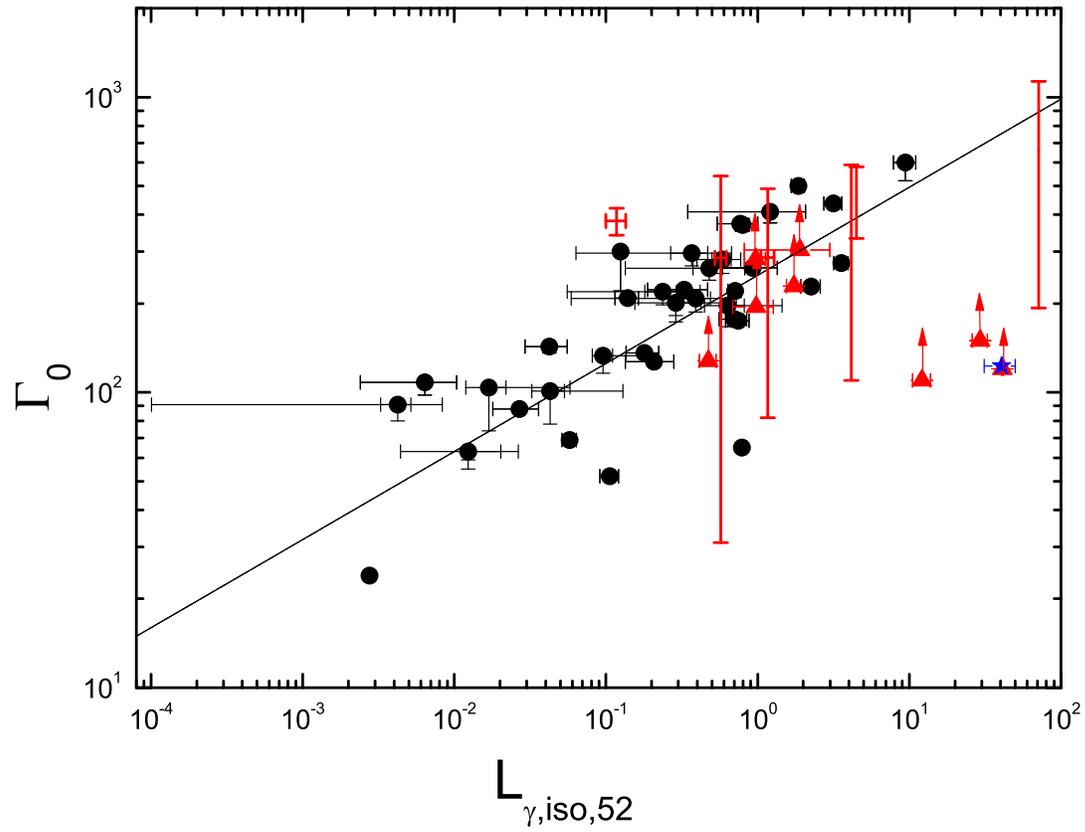}}
 \caption{The initial Lorentz factor $\Gamma_{\rm 0}$ vs. the isotropic equivalent $\gamma$-ray luminosity $L_{\rm \gamma,iso}$. The best fitting line is $\Gamma_{\rm 0} \simeq 249 L_{\rm \gamma,iso,52}^{0.30}$ with correlation coefficient $\zeta=0.79$. The notations are the same as in Fig. 1.}
\end{center}
\label{fig:GL}
\end{figure}

%--------------------- Table --------------------------
\small
\renewcommand{\arraystretch}{0.6}
\renewcommand{\thefootnote}{\fnsymbol{footnote}}
\begin{deluxetable}{lrrrrrr}
\tablewidth{0pt}
\tablecaption{The quantities of the GRBs in our sample. Main references are listed in the last column. Others are marked directly after the quantities. \label{Tab:publ-works}}
\tablehead{
\colhead{GRB}                   & \colhead{$z$}      &
\colhead{$\Gamma_{0}$}          & \colhead{$E_{\rm \gamma,iso, 52}$}  &
\colhead{$T_{\rm 90}$}          & \colhead{$L_{\rm \gamma,iso, 52}$}    &
\colhead{Refs} }
\startdata
990123 & $1.61^{[1,2]}$ & $600\pm79.6$ & $229\pm37$ & $63.3\pm0.3$ & $9.44\pm1.57$ & a \\
021211 & $1.006^{[3]}$ & $>195$ & $1.12\pm0.08$ & $2.3\pm0.52^{[44]}$ & $0.98\pm0.29$ & b \\
040924 & $0.858^{[4]}$ & $(82,490)$ & 1.5 & $2.39\pm0.24^{[45]}$ & $1.17\pm0.12$ & b,c \\
050401 & $2.9^{[5]}$ & $(110,590)$ & $35^{[5]}$ & $33^{[5]}$ & $4.14$ & c \\
050525A & $0.606^{[6]}$ & $>229$ & $9.54\pm0.52$ & $8.8\pm0.5^{[46]}$ & $1.74\pm0.19$ & b \\
050730 & $3.97^{[7]}$ & $201^{+28}_{-19}$ & $9^{+8}_{-3}$ & $155\pm20^{[47]}$ & $0.29^{+0.29}_{-0.13}$ & b \\
050801 & $1.56^{[8]}$ & $(341,420)$ & $0.916^{[8]}$ & $20\pm3^{[48]}$ & $0.12\pm0.018$ & b,c \\
050820A & $2.615^{[9]}$ & $282^{+29}_{-14}$ & $97^{+31}_{-14}$ & $\sim600^{[49]}$ & $0.58^{+0.19}_{-0.08}$ & b \\
050922C & $2.198^{[10]}$ & $274$ & 3.7 & $4.54^{[50]}$ & $3.56\pm0.39$ & b \\
060210 & $3.91^{[11]}$ & $264\pm4$ & $41.5\pm5.7$ & $220\pm70^{[51]}$ & $0.93\pm0.42$ & b \\
060418 & $1.49^{[12,13]}$ & $263^{+23}_{-7}$ & $10^{+7}_{-2}$ & $52\pm1^{[12]}$ & $0.48^{+0.34}_{-0.1}$ & b \\
060605 & $3.8^{[14]}$ & $197^{+30}_{-6}$ & $2.5^{+3.1}_{-0.6}$ & $19\pm1^{[14]}$ & $0.63^{+0.82}_{-0.18}$ & b \\
060607A & $3.082^{[12,15]}$ & $296^{+28}_{-8}$ & $9^{+7}_{-2}$ & $100\pm5^{[52]}$ & $0.37^{+0.3}_{-0.1}$ & b \\
060614 & $0.125^{[16]}$ & 24 & 0.25 & $102^{[53]}$ & $0.0028$ & c \\
060904B & $0.703^{[17]}$ & $108\pm10$ & $0.72\pm0.43$ & $192\pm5^{[54]}$ & $0.0064\pm0.004$ & b \\
060908 & $2.43^{[18]}$ & $>304$ & $10.7\pm5.94$ & $19.3\pm0.3^{[55]}$ & $1.9\pm1.09$ & b \\
061007 & $1.262^{[19]}$ & $436\pm3$ & $104.65\pm6.94$ & $75\pm5^{[56]}$ & $3.16\pm0.42$ & b \\
061121 & $1.314^{[20]}$ & $175\pm2$ & $26.1\pm3$ & $81\pm5^{[15]}$ & $0.75\pm0.13$ & d \\
070110 & $2.352^{[21]}$ & $127\pm4$ & $5.5\pm1.5$ & $89\pm7^{[57]}$ & $0.21\pm0.073$ & d \\
070318 & $0.84^{[22]}$ & $143\pm7$ & $1.45\pm0.38$ & $63\pm3^{[58]}$ & $0.042\pm0.013$ & b \\
070411 & $2.954^{[23]}$ & $208^{+21}_{-5}$ & $10^{+8}_{-2}$ & $101\pm5^{[59]}$ & $0.39^{+0.33}_{-0.098}$ & b \\
070419A & $0.97^{[24]}$ & $91^{+11}_{-3}$ & $0.24^{+0.23}_{-0.05}$ & $112\pm2^{[60]}$ & $0.0042^{+0.0041}_{-0.00095}$ & b \\
071003 & $1.1^{[25]}$ & $>283$ & $68.4\pm10.4$ & $148\pm1^{[25]}$ & $0.97\pm0.15$ & b \\
071010A & $0.98^{[26]}$ & $101^{+23}_{-3}$ & $0.13^{+0.24}_{-0.01}$ & $6\pm1^{[61]}$ & $0.043^{+0.086}_{-0.01}$ & b \\
071010B & $0.947^{[27]}$ & $209\pm4$ & $2.55\pm0.41$ & $35.74\pm0.5^{[62]}$ & $0.14\pm0.024$ & b \\
071031 & $2.692^{[28]}$ & $133^{+17}_{-3}$ & $3.9^{+4.1}_{-0.6}$ & $150.49^{[50]}$ & $0.096^{+0.1}_{-0.015}$ & b \\
080129 & $4.394^{[29,30]}$ & 65 & 7 & $48^{[29]}$ & $0.79$ & a \\
080319B & $0.937^{[31]}$ & $(332,580)$ & $132$ & $57^{[63]}$ & $4.49$ & b,c \\
080319C & $1.95^{[32]}$ & $228\pm5$ & $22.55\pm3.35$ & $29.55^{[50]}$ & $2.25\pm0.33$ & b \\
080330 & $1.51^{[33]}$ & $104^{+30}_{-2}$ & $0.41^{+0.94}_{-0.06}$ & $61\pm9^{[64]}$ & $0.017^{+0.041}_{-0.005}$ & b \\
080413B & $1.1^{[34]}$ & $>128$ & 1.8 & $8.0\pm1.0$ & $0.47\pm0.059$ & e \\
080603A & $1.688^{[35]}$ & 88 & $1.5\pm0.5$ & $150^{[65]}$ & $0.027\pm0.009$ & a \\
080710 & $0.845^{[36]}$ & $63^{+8}_{-4}$ & $0.8^{+0.8}_{-0.4}$ & $120\pm17^{[66]}$ & $0.012^{+0.014}_{-0.008}$ & b \\
080810 & $3.35^{[37]}$ & $409\pm34$ & $30\pm20$ & $108\pm5^{[67]}$ & $1.21\pm0.86$ & b \\
080916C & $4.35^{[38]}$ & $(193,1130)$ & 880 & $66^{[38]}$ & 71.33 & c,f \\
081203A & $2.1^{[39]}$ & $219^{+21}_{-6}$ & $17^{+13}_{-4}$ & $223^{[50]}$ & $0.24^{+0.18}_{-0.056}$ & b \\
090313 & $3.375^{[40]}$ & 136 & 3.2 & $78\pm19^{[68]}$ & $0.18\pm0.044$ & a \\
090323 & $3.568^{[41]}$ & $>110$ & $399\pm53$ & $150^{[44]}$ & $12.15\pm1.61$ & f \\
090328A & $0.736^{[41]}$ & $(31,540)$ & $23\pm2$ & $70^{[44]}$ & $0.57\pm0.05$ & c,f \\
090424 & $0.544^{[42]}$ & $300\pm79$ & $4$ & $49.47^{[50]}$ & $0.12$ & c \\
090510 & $0.903^{[31]}$ & 123$^\dagger$\footnote[0]{$^\dagger$We note this special short burst GRB 090510 was reported with $\Gamma_0 > 1200$ \citep{Ackermann2010}, which uses method B with one-zone assumption. However, it is inconsistent with other methods as shown in \citet{Zou2010}. Here we adopt the $\Gamma_0$ by method A, and it is consistent with the modified method B (two-zone assumption).} & 6.4 & $0.3\pm0.07^{[69]}$ & $40.3\pm9.47$ & c \\
090812 & $2.452^{[43]}$ & 501 & $40.3\pm4$ & $75.09^{[50]}$ & $1.85\pm0.18$ & d \\
090902B & $1.8229^{[41]}$ & $>120$ & $320\pm4^{[41]}$ & $21.9^{[70]}$ & $41.25\pm0.52$ & f \\
090926A & $2.1062^{[41]}$ & $>150$ & $189\pm3^{[41]}$ & $20\pm2^{[71]}$ & $29.35\pm3.4$ & f \\
091024 & $1.092^{[43]}$ & 69 & $28\pm3$ & $1020^{[72]}$ & $0.057\pm0.0062$ & d \\
091029 & $2.752^{[43]}$ & 221 & $7.4\pm0.74$ & $39.18^{[50]}$ & $0.71\pm0.071$ & d \\
100621A & $0.542^{[43]}$ & 52 & $4.37\pm0.5$ & $63.6\pm1.7^{[73]}$ & $0.11\pm0.015$ & d \\
100728B & $2.106^{[43]}$ & 373 & $3\pm0.3$ & $12.1\pm2.4^{[74]}$ & $0.77\pm0.23$ & d \\
100906A & $1.727^{[43]}$ & 369 & $33.4\pm3$ & $114.4\pm1.6^{[75]}$ & $0.8\pm0.083$ & d \\
110205A & $2.22^{[43]}$ & 177 & $56\pm6$ & $257\pm25^{[76]}$ & $0.7\pm0.14$ & d \\
110213A & $1.46^{[43]}$ & 223 & $6.4\pm0.6$ & $48\pm16^{[77]}$ & $0.33\pm0.14$ & d \\
\enddata

\tablerefs{[1] \citet{Akerlof1999}; [2] \citet{Galama1999}; [3] \citet{Vreeswijk2006};  [4] \citet{Wiersema2005}; [5] \citet{De Pasquale2006}; [6] \citet{Foley2005}; [7] \citet{Rol2005}; [8] \citet{De Pasquale2007}; [9] \citet{Ledoux2005}; [10] \citet{Jakobsson2005}; [11] \citet{Cucchiara2006}; [12] \citet{Molinari2007}; [13] \citet{Falcone2006}; [14] \citet{Ferrero2009}; [15] \citet{Page2007}; [16] \citet{Mundell2007}; [17] \citet{Fugazza2006}; [18] \citet{Rol2006}; [19] \citet{Jakobsson2007a}; [20] \citet{Bloom2006}; [21] \citet{Jaunsen2007}; [22] \citet{Chen2007}; [23] \citet{Jakobsson2007a}; [24] \citet{Cenko2007b}; [25] \citet{Perley2008a}; [26] \citet{Prochaska2007}; [27] \citet{Cenko2007a}; [28] \citet{Ledoux2007}; [29] \citet{Greiner2009}; [30] \citet{Stratta2009}; [31] \citet{Rau2009}; [32] \citet{Wiersema2008}; [33] \citet{Cucchiara2008}; [34] \citet{Fynbo2009}; [35] \citet{Guidorzi2009b}; [36] \citet{Perley2008}; [37] \citet{Prochaska2008}; [38] \citet{Abdo2009a}; [39] \citet{Landsman2008}; [40] \citet{Melandri2010}; [41] \citet{Cenko2011}; [42] \citet{Wiersema2005}; [43] \citet{Ghirlanda2011}; [44] \citet{Crew2003}; [45] \citet{Donaghy2006};  [46] \citet{Blustin2006}; [47] \citet{Pandey2006}; [48] \citet{Rykoff2006}; [49] \citet{Cenko2006}; [50] \citet{Sakamoto2011}; [51] \citet{Curran2007}; [52] \citet{Ziaeepour2008}; [53] \citet{Gehrels2006}; [54] \citet{Klotz2008}; [55] \citet{Covino2010}; [56] \citet{Schady2007}; [57] \citet{Troja2007}; [58] \citet{Chester2008}; [59] \citet{Ferrero2008}; [60] \citet{Melandri2009}; [61] \citet{Kong2010}; [62] \citet{Wang2008}; [63] \citet{Racusin2008}; [64] \citet{Yuan2008}; [65] \citet{Martin-Carrillo2008}; [66] \citet{Kruhler2009}; [67] \citet{Page2009}; [68] \citet{de2010}; [69] \citet{De Pasquale2010}; [70] \citet{Abdo2009c}; [71] \citet{Swenson2010}; [72] \citet{Gruber2011}; [73] \citet{Ukwatta2010}; [74] \citet{Barthelmy2010a}; [75] \citet{Barthelmy2010b}; [76] \citet{Markwardt2011}; [77] \citet{Barthelmy2011}; (a) \citet{Melandri2010}; (b) \citet{Liang2010}; (c) \citet{Zou2010}; (d) \citet{Ghirlanda2011}; (e) \citet{Filgas2011}; (f) \citet{Zou2011}.}

\end{deluxetable}

\end{document}